\documentclass{aa}
\usepackage{psfig}
\usepackage{epsfig}
\usepackage{graphicx}

\def \hcm {\hbox {\ifmmode $ cm$^{-2}\else cm$^{-2}$\fi}}
\def \arcmin {\hbox{$^\prime$}}
\def \arcsec {\hbox{$^{\prime\prime}$}}

\def\approxgt{\mathrel{\hbox{\rlap{\lower.55ex \hbox {$\sim$}}
        \kern-.3em \raise.4ex \hbox{$>$}}}}
\def\approxlt{\mathrel{\hbox{\rlap{\lower.55ex \hbox {$\sim$}}
        \kern-.3em \raise.4ex \hbox{$<$}}}}

\begin{document}

\title{XMM-Newton spectral analysis of the Pulsar Wind Nebula 
within the composite SNR \object{G0.9+0.1}}

\author{D. Porquet\inst{1} 
        \and A. Decourchelle\inst{1} 
        \and R. S. Warwick\inst{2}
       }
\offprints{Delphine Porquet (dporquet@cea.fr)}

\institute{Service d'Astrophysique, Orme des Merisiers, 
CE-Saclay, 91191 Gif-sur-Yvette, Cedex, France 
\and Department of Physics and Astronomy, University of Leicester, 
Leicester LE1 7RH, UK }
\date{Received ...; Accepted ...  }

\markboth{X--rays from G0.9+0.1}{X--rays from G0.9+0.1}

\abstract{
 We present a study of the composite supernova remnant G0.9+0.1 based on
observations by {\sl XMM-Newton}. The EPIC spectrum shows 
  diffuse X-ray emission from the region corresponding to the radio shell. 
The X--ray spectrum of the whole 
Pulsar Wind Nebula is well fitted by an absorbed 
power--law model with a photon index $\Gamma \sim$ 1.9 
and a 2--10 keV luminosity of about 
$6.5 \times 10^{34}~{\rm d}_{\mathrm{10}}^{2} $ erg s$^{-1}$  
(d$_{10}$ is the distance in units of 10\,kpc).
However, there is a clear softening of the X-ray spectrum
with distance from the core, which is most probably related 
to the finite lifetime of the synchrotron emitting electrons.
This is fully consistent with the plerionic 
interpretation of the Pulsar Wind Nebula, in
which an embedded pulsar injects energetic electrons into its 
surrounding region.
At smaller scales, the eastern part of the arc-like feature, which was
first revealed by {\sl Chandra} observations, shows indications of a hard X-ray
spectrum with a corresponding small photon index ($\Gamma$=1.0$\pm$0.7), while the 
western part presents a significantly softer spectrum ($\Gamma$=3.2$\pm$0.7). 
A possible explanation for this feature is fast rotation and subsequent Doppler boosting 
of electrons: the eastern part of the torus has a velocity component pointing 
towards the observer, while the western part has a velocity component in the 
opposite direction pointing away from the observer. 
}
\titlerunning{G0.9+0.1 observed by {\sl XMM-Newton}.}
\authorrunning{Porquet, Decourchelle, Warwick}
\maketitle

\section{Introduction}

\begin{figure*}[!t]
\includegraphics[angle=0,width=\textwidth]{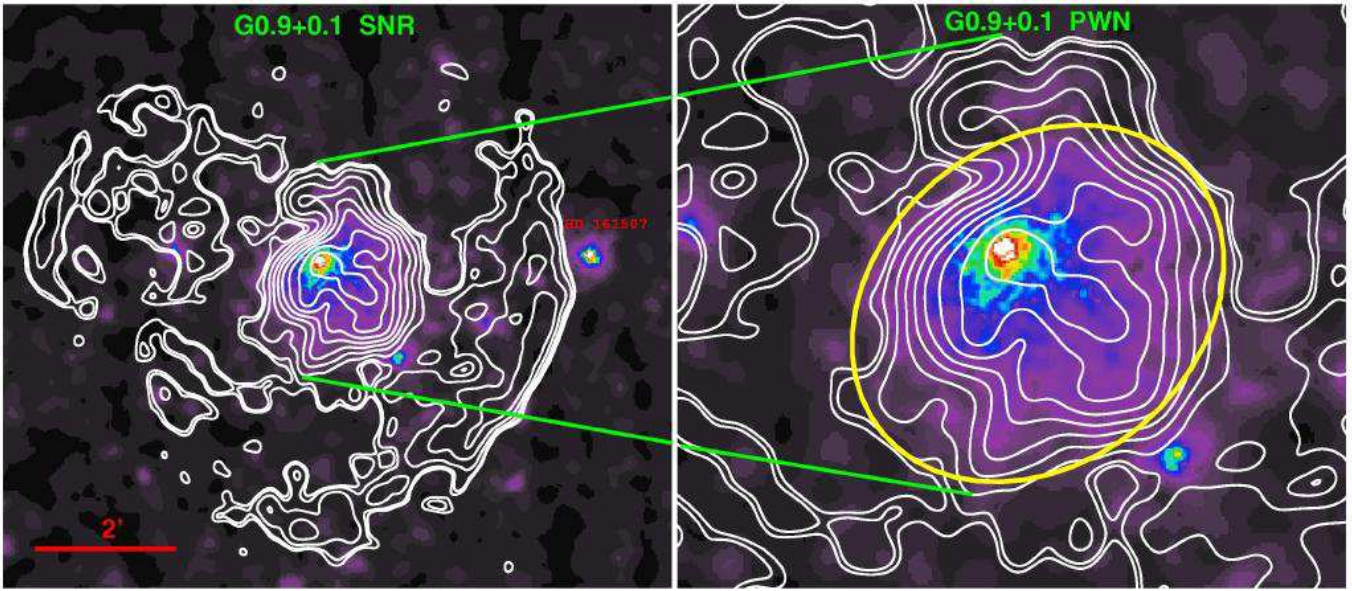}
      \caption[]{{\sl XMM-Newton} EPIC image of SNR G0.9+0.1 and its PWN in 
the energy band 1.5--12 keV  with an adaptative smoothing filter
with a signal-to-noise requirement of 5 and gaussian smooth of 
10\arcsec. 
The  {\sl VLA} radio contours at 1.5 GHz (20\,cm) are superimposed in white. 
{\it Left panel}: overall remnant. 
{\it Right panel}: pulsar wind nebula (PWN). The yellow ellipse 
represents the region taken for spectra analysis of the overall PWN.
}
   \label{fig:radio}
\end{figure*}

The radio source G0.9+0.1 is the only composite SNR, listed in the Green 
catalog (Green 2001), which lies in the general direction of the 
Galactic Center. According to Helfand \& Becker (\cite{Helfand87}), 
the radio morphology of G0.9+0.1 is characterized by a bright centrally 
condensed synchrotron nebula with relatively flat energy spectral index 
($\sim 2'$ diameter, $\alpha$=+0.12 with F$_{\rm \nu}\propto \nu^{-\alpha}$) 
and a radio shell with a 
steeper radio spectrum  ($\sim 8'$ diameter; $\alpha$=+0.77).
The synchrotron core is powered by the loss of rotational energy from a 
central pulsar; hereafter we refer to this core as the
Pulsar Wind Nebula (PWN) in G0.9+0.1.

Owing to the  high interstellar absorption
(${\cal{N}}_{\rm H}\sim10^{23}$~\hcm),  G0.9+0.1  was barely detected by the 
{\sl Einstein} Observatory (Helfand \& Becker \cite{Helfand87}). However, a 
much more convincing detection was obtained at higher energies with the 
{\sl BeppoSAX} satellite (Mereghetti et al. \cite{Mereghetti98}). 
Later on, Sidoli et al. (\cite{Sidoli2000}) interpreted this hard X-ray 
emission as non-thermal in origin. The small angular extent of the X-ray 
source (radius $\sim$ 1\,\arcmin), combined with an estimated age of the 
remnant of a few thousand years, is further evidence that the central radio 
core is powered by a young pulsar ($\sim$ 2,700\,yr; 
Sidoli et al. \cite{Sidoli2000}). The SNR shell remains unobserved in X-rays,  
probably due to the suppression of its soft X-ray flux by the 
line-of-sight absorption.\\
\indent Recently, G0.9+0.1 was observed with {\sl Chandra} (Gaensler et al. 
\cite{Gaensler2001}). The high angular resolution of {\sl Chandra} provided
an unprecedented view of the X-ray morphology of this SNR and identified 
a semi-circular arc-like feature, a jet-like feature and a very bright and 
small central emission region close to a very faint unresolved source 
(\object{CXOU J174722.8-280915}). The latter is inferred to be the pulsar 
itself (Gaensler et al. \cite{Gaensler2001}). 
This X-ray morphology is similar to other PWNe  
powered by young pulsars, which display jets combined with a torus structure
 (\object{Crab}: Brinkmann et al. \cite{Brinkmann85}, 
Weisskopf et al. \cite{Weisskopf2000}; 
\object{Vela}: Pavlov et al. \cite{Pavlov2000}).
As mentioned by Gaensler et al. (\cite{Gaensler2001}), 
the limited counting statistics of the 35\,ks 
{\sl Chandra} observation did not allow them to carry out 
a detailed spectral study of the various structures of the PWN,
as ideally is required for the testing of current models. \\

\indent Here we present the results from an observation of the composite 
SNR G0.9+0.1 made with {\sl XMM-Newton}. We focus on the 
 spectral analysis of different regions of the remnant
on both relatively large and small spatial scales.

\section{Observations and data analysis}
 
G0.9+0.1 was observed by {\sl XMM-Newton} on-axis 
on September 23--24, 2000 ($\sim$17.2\,ks and $\sim$12\,ks 
for the MOS and PN cameras, respectively). 
The EPIC-MOS cameras were operated in the standard full-frame mode 
and the EPIC-PN camera in the extended full frame mode. Each camera
was operated with the medium filter deployed.
 
Using the {\sc XMM Science Analysis Software} (SAS version 5.2),
the recorded events were screened by rejecting the high background periods
which occasionally arise due to an intense incident flux of soft protons.
After this data cleaning, the useful observing times are 
respectively for MOS1 and MOS2 about 15.5\,ks and 15.4 ks, 
and 10\,ks for PN. X-ray events corresponding to pattern 0--12 for the two 
MOS cameras  were used, whereas for the PN only pattern 0--4 
(single and double pixel events) were accepted. 
The astrometry of this observation was substantially improved using as 
reference the bright foreground F3V star \object{HD 161507},  
which is an eclipsing binary of Algol type, detected 
at the western side of G0.9+0.1 (see Fig.~\ref{fig:radio}: 
{\it left panel}).\\ 
For the imaging analysis we have summed up the MOS and PN data. 
 We have used for the adaptative smoothing the tool {\sc asmooth} 
from the SAS.\\  
Our spectral analysis combines fits of the MOS and PN data (except in $\S$3.3)
employing the following camera response matrices: 
m1$\_$medv9q20t5r6$\_$all$\_$15.rsp, 
m2$\_$medv9q20t5r6$\_$all$\_$15.rsp,
epn$\_$ef20$\_$sdY9$\_$medium.rsp. 
We subtract from the source and the local background in our pointing,  
complementary data from a blank-field observation (kindly provided by 
David Lumb) in order to take into account the particle background 
(Majerowicz et al. \cite{Maj2002}).
The normalization between our pointing and the blank-field files  
is determined by their count ratios at [10--12]\,keV 
for the MOS and at [12--14]\,keV for the PN. 
For the vignetting, we applied the weighting method 
described by Arnaud et al. (\cite{Arnaud2001}).
 Our spectra were binned to 3$\sigma$ before background subtraction, 
except in $\S$\ref{sec:shell}.
{\sc xspec v11.1.0} was used for the spectral analysis.
All subsequent errors are quoted at 90$\%$ confidence. 
Abundances are those of Anders \& Grevesse (\cite{Anders89}).

\section{The composite SNR}\label{sec:SNR}

The radio data of SNR G0.9+0.1 clearly show a radio shell and 
a very bright centrally condensed synchrotron nebula 
(Helfand \& Becker \cite{Helfand87}).
Within the PWN, two central (west and east) peaks 
are present at 20\,cm, while at 6\,cm only the west peak is visible
(Fig.1~a,b in Helfand \& Becker \cite{Helfand87}).
Up to now, there is no known  X-ray counterpart to the radio shell, 
while the PWN is clearly detected above about 2--3\,keV 
(Mereghetti et al. \cite{Mereghetti98}, Gaensler et al. 
\cite{Gaensler2001}). \\
Figure~\ref{fig:radio} shows the {\sl XMM-Newton} EPIC images of the composite 
SNR G0.9+0.1 in the energy band 1.5--12\,keV with the radio contours overlaid. 
We use the radio data initially from Helfand \& Becker (\cite{Helfand87}) 
which have been re-analyzed by Gaensler et al. (\cite{Gaensler2001}). 
The spatial resolutions are respectively 8$\arcsec$ for {\sl XMM-Newton}
and 40$\arcsec$ for the {\sl VLA} data. 
Several X-ray bright point sources are observed 
within the field such as HD\,161507 (a F3V star) 
and a bright X-ray source located to the south-west of G0.9+0.1 which 
appears to correspond to the 2MASS source
\object{2MASS1747178-281025} (according to the cross-correlation results
from {\sl XMM-Newton} EPIC pipeline processing). 
 A more focused spectral study of the radio shell region brings for the first time  
 an indication of diffuse X-ray emission, which will be addressed in more detail
below. Furthermore, the large scale X-ray morphology of the PWN (R$\sim$1')
is in good agreement with the {\sl VLA} radio contours 
at 20\,cm  (Fig~\ref{fig:radio}: {\it right panel}).

\subsection{The SNR shell}\label{sec:shell}

\begin{figure}[!ht]
\hspace{-2.5cm}
\hspace*{2.4cm}\psfig{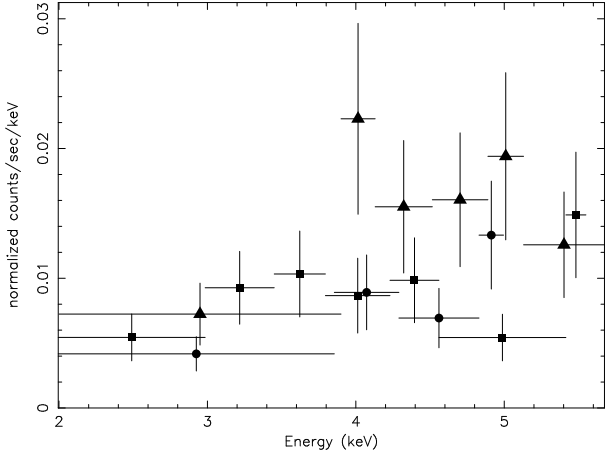}
\hspace{-2.5cm}
      \caption[]{XMM-Newton spectra
 (MOS1: filled circles, MOS2: filled squares, and PN: filled triangles)
 of the shell region (R$\sim$4', excluding the PWN) of G0.9+0.1. }
   \label{fig:spectrumshell}
\end{figure}
We extracted spectra of the region corresponding 
to the radio shell (R$\sim$ 4'), excluding the PWN 
(see ellipse in Fig.~\ref{fig:radio}: {\it right panel}). 
We adopt a 3$\sigma$ binning above background. Since the spectrum is strongly 
absorbed below 2 keV, due to the interstellar medium in the Galactic Center 
region, we focus on the energy band 2--10 keV. The background 
 taken for the shell analysis 
is an annulus centered on the PWN 
within radii  4.4\arcmin~ and 8.8\arcmin, 
excluding the bright sources.
The background subtracted spectra 
 shown in Figure~\ref{fig:spectrumshell},
  clearly reveal a signal above zero level 
which can be interpreted as the first detection of X-ray 
emission in this region. 
  We fitted the combined MOS and PN spectra
with both a non-thermal model ({\sc power-law}) 
and a thermal model ({\sc mekal}) with ${\cal{N}}_{\rm H}$ fixed
at 1.39$\times$10$^{23}$\,cm$^{-2}$ as determined 
for the PWN (see $\S$\ref{sec:PWN}). 
We find for the {\sc mekal} model 
 kT=3.4$^{+5.3}_{-1.2}$\,keV ($\chi^2$/d.o.f.= 17.3/15), 
and for the power-law model $\Gamma$= 2.6$^{+0.7}_{-0.8}$ 
($\chi^2$/d.o.f.= 17.4/15). 
The inferred temperature is relatively high for a several thousand year old
SNR.
\begin{figure*}[!Ht]
\includegraphics[angle=0,width=\textwidth]{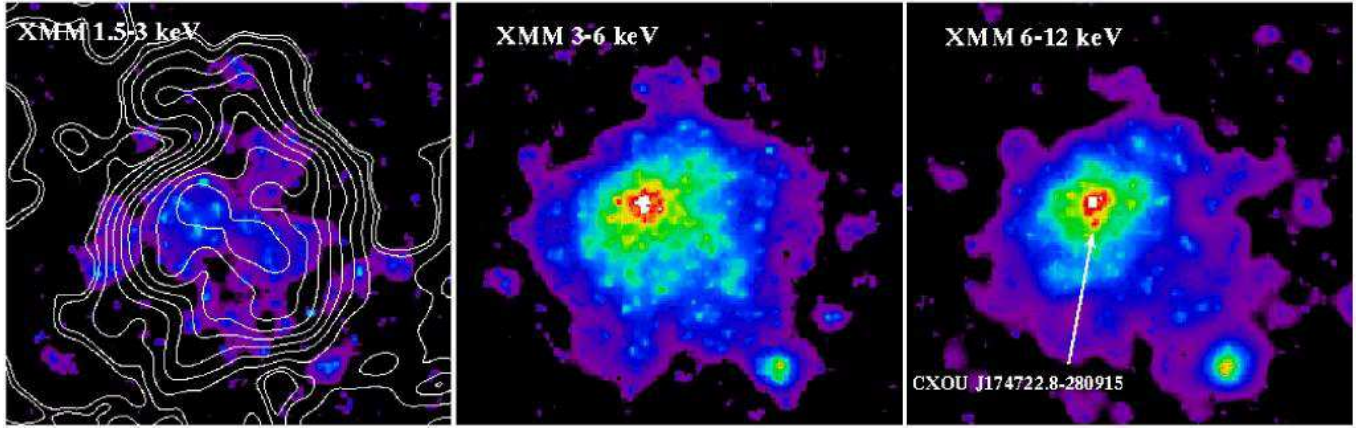}
\caption[]{{\sl XMM-Newton} EPIC image of the PWN in different energy 
bands, obtained with adaptative smoothing with a signal-to-noise
requirement of 5 and a gaussian smooth of 20\arcsec. 
The {\sl VLA} radio contours at 1.5 GHz (20\,cm) are superimposed in white. }
   \label{fig:bandes}
\end{figure*}
Both models give respectively an unabsorbed flux (2--10\,keV), 
of 2.2$^{+4.1}_{-1.5}\times$10$^{-12}$ erg\,cm$^{-2}$\,s$^{-1}$, 
and 2.3$^{+1.4}_{-0.9}\times$10$^{-12}$ erg\,cm$^{-2}$\,s$^{-1}$.
The inferred fluxes are in good agreement with the upper limit
 quoted by Sidoli et al. (\cite{Sidoli2000}),
i.e. 3$\times$10$^{-12}$ erg\,cm$^{-2}$\,s$^{-1}$. 
We also carried out a spectral analysis 
(fixing ${\cal{N}}_{\rm H}$ at 1.39$\times$10$^{23}$\,cm$^{-2}$)
of the brightest western region of the radio shell 
(the statistic of the eastern part is too poor to allow
a spectral analysis). 
Again, this region is dominated by the local diffuse emission 
of the GC region and thus requires appropriate background subtraction, 
which we extract from an adjacent field. 
The corresponding fit results for the temperature
is  kT$=$2.7$^{+17.6}_{-1.2}$\,keV ($\chi^2$/d.o.f.= 3.5/5), 
and is thus not well constrained.
 A power law model yields $\Gamma$=2.9$^{+1.2}_{-1.3}$ 
($\chi^2$/d.o.f.= 3.3/5).\\
\indent We extracted an image in the energy band 3.9--5.8\,keV, 
i.e. in the spectral interval with maximum emission.
However, no distinct borderline is detected between the X-ray shell
 and the diffuse medium located in the GC region.

\subsection{The PWN}\label{sec:PWN}

\begin{figure}[!ht]
\hspace{-2.5cm}
\hspace*{2.4cm}\psfig{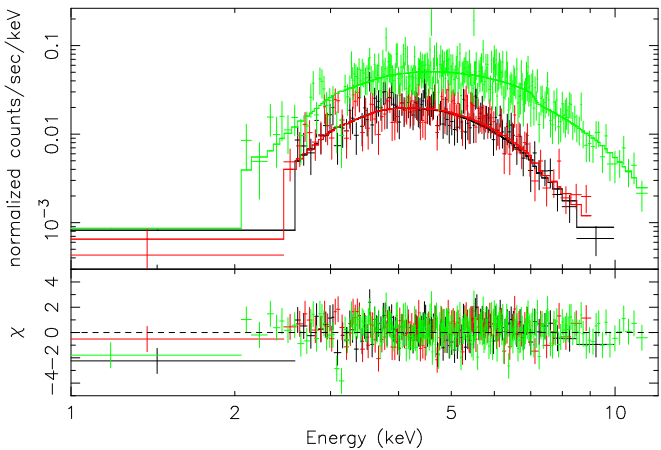}
\hspace{-2.5cm}
      \caption[]{Combined fit (MOS1 in black, MOS2 in red, and PN in green)
 of the PWN of G0.9+0.1 with an absorbed power law ({\sc tbabs*po}). 
See Table~\ref{table:table1} for the inferred parameters. 
}
   \label{fig:spectrumPWN}
\end{figure}

We observe that the  core 
of the PWN is surrounded by extended diffuse emission 
(Fig.~\ref{fig:radio}: {\it right panel}). 
The position of the X-ray core matches the eastern radio peak 
while there is no bright X-ray counterpart to the western radio peak.
Figure~\ref{fig:bandes} displays the PWN in different energy bands, 
namely 1.5--3\,keV, 3--6\,keV, and 6--12\,keV. 
In the energy band 1.5--3\,keV, there is little evidence
for a dominant core and the radio and X-ray morphologies are similar 
 (see Fig.~\ref{fig:bandes}: {\it left panel}). 
The core of the PWN is only observable above 
$\sim 3$\,keV (Fig~\ref{fig:bandes}: {\it middle and right panels}). 
 At high energy (6--12\,keV) the symmetry axis is analogous
 to the one of {\sl Chandra} data, the western radio lobe is very toned down, 
 and the faint X-ray source CXOU J174722.8-280915 is apparent  
(Fig~\ref{fig:bandes}: {\it right panel}).\\  
For comparison with previous studies, we fit the spectrum of the overall 
PWN. The region considered is defined by the ellipse displayed 
in Figure~\ref{fig:radio} ({\it right panel}). 
Our spectra were binned to 3$\sigma$ before background subtraction.
We defined a local background corresponding 
to the central MOS\,1 CCD region, 
excluding the bright point sources and the central part 
within a radius of $2.6'$. 
As shown in Figure~\ref{fig:spectrumPWN}, 
 the corresponding spectra extend up to almost 12\,keV for the 
PN data, and to about 9\,keV in the MOS data, allowing strong constraints 
to be placed on the slope of the continuum.  
The data are well fitted by either an absorbed power-law  
or a thermal bremsstrahlung model. 
For the photo-electric absorption (${\cal{N}}_{H}$), we use first the 
cross-sections of Morrison \& McCammon (\cite{Morrison83}, {\sc wabs}). 
The best-fit parameters are reported in Table~\ref{table:table1}. 
We also applied the same spectral fitting to spectra with larger binsizes 
(higher signal to noise) and obtained fit results, which were in very good
agreement with the values quoted here.
The best-fit parameters relating to the power-law are compatible,
 within the error bars, with the {\sl BeppoSAX} and {\sl Chandra} values 
(Sidoli et al. \cite{Sidoli2000}: 
${\cal{N}}_{H}$=1.09$^{+0.24}_{-0.21}\times$10$^{23}$\,cm$^{-2}$, 
$\Gamma$=1.95$^{+0.33}_{-0.30}$; 
Gaensler et al. \cite{Gaensler2001}: 
${\cal{N}}_{H}$=1.6$\pm 0.02 \times$10$^{23}$\,cm$^{-2}$, 
$\Gamma$=2.4$\pm0.4$), 
though more tightly constrained. 

\begin{table}[!hb]
\begin{center}
\caption[]{
Results of the spectral fits for the X-ray PWN. 
The assumed interstellar photo-electric absorption cross-sections are  
from Morrison \& McCammon (\cite{Morrison83}; {\sc wabs} in {\sc XSPEC}) 
or from Wilms et al. (\cite{Wilms2000}; {\sc tbabs}).
Uncertainties are quoted at 90$\%$ confidence. 
The unabsorbed fluxes (2--10\,keV) are expressed in $10^{-12}$ erg\,cm$^{-2}$\,s$^{-1}$. 
}
\label{table:table1}
\begin{tabular}{ccccc}
\hline
\hline
\noalign {\smallskip}
Model          & ${\cal N}_{H}$        & {\small $\Gamma$ or $kT$} & $\chi^2$  & F$^{\mathrm \tiny unabs}_{2-10}$ \\
               & {\tiny ($10^{23}$\,cm$^{-2}$)}&      \,\,\,\,\,\,\,\,{\tiny (keV)}&  {\tiny (446 d.o.f.)}  &\\
\hline
wabs*pow       & 1.47$^{+0.14}_{-0.13}$      & 1.99$^{+0.19}_{-0.18}$   &   {\small 379.6} & 5.78   \\
wabs*brems    & 1.33$^{+0.11}_{-0.10}$      & 10.7$^{+4.2}_{-2.3}$    &   {\small 379.1}  & 5.16  \\
\hline
tbabs*pow      & 1.39$^{+0.13}_{-0.12}$      & 1.93$^{+0.18}_{-0.18}$   &   {\small 378.5}    & 5.73\\
 tbabs*brems    & 1.27$^{+0.10}_{-0.10}$      & 11.7$^{+4.9}_{-2.7}$    &   {\small 378.0}   & 5.17\\
\hline
\noalign {\smallskip}                       
\hline
\end{tabular}
\end{center}
\end{table}

In the following, we use the updated cross-sections for X-ray absorption by 
the interstellar medium ({\sc tbabs} in {\sc xspec}) from Wilms et al. (\cite{Wilms2000}).  
The best-fit parameters although in good agreement with the previous ones
 (see Table~\ref{table:table1}), give systematically slightly lower 
${\cal{N}}_{H}$ values using {\sc tbabs} absorption model 
than {\sc wabs} absorption model.
We obtain for the power-law model a luminosity in the 2--10\,keV energy
band of $6.5 \times 10^{34}~{\rm d}_{\mathrm{10}}^{2} $ erg s$^{-1}$ 
(d$_{10}$ is the distance in units of 10\,kpc). 


\subsection{CXOU J174722.8-280915}

CXOU J174722.8-280915  
is a point source located about 8\arcsec~below the X-ray bright 
core  of the PWN in the {\sl Chandra} observations and has
been identified as the pulsar itself by Gaensler et al. (\cite{Gaensler2001}). 
In our {\sl XMM-Newton} data, we are able to pick out this point 
source only above 6\,keV (Fig~\ref{fig:bandes}: {\it right panel}).\\
  
We extracted a spectrum from a circular region 
(R=5\arcsec) centred on the position of the point source 
(Fig~\ref{fig:bandes}: {\it right panel}).
 For the background we took a region near the 
 source avoiding any strong contamination from the core emission
(due to the point spread function of the EPIC cameras).
We used only the spectra from the MOS cameras 
which benefit from the best mirrors and 
pixel sampling. 
As in Gaensler et al. (\cite{Gaensler2001}), 
we explore the possibility that CXOU J174722.8-280915 corresponds
to a young
central pulsar. Its X-ray emission is then expected to be due 
to either modified black-body emission from the neutron star surface 
or non-thermal emission from the neutron star magnetosphere
(e.g. Becker \& Tr{\"u}mper \cite{Becker97}). 
As calculated by van Ripert, Link \& Epstein (\cite{vanRiper95}),
 a neutron star with an age between 10$^{3}$ and 10$^{4}$ years 
should have a surface temperature of kT$\sim$0.1\,keV.
Fixing $\cal{N}_{\rm H}$ at 1.39$\times$10$^{23}$\,cm$^{-2}$ 
and the temperature at 0.1\,keV, we obtain a very bad fit 
with $\chi^2_{\rm red}$=2.4 (5 d.o.f.). 
Letting kT or  $\cal{N}_{\rm H}$ free does 
not improve the fit. In contrast when we assume a non-thermal model
 (as in Gaensler et al. \cite{Gaensler2001}), with  
a photon index of $\Gamma$=1.5 (typical for magnetospheric
emission, Becker \& Tr{\"u}mper \cite{Becker97}) 
and $\cal{N}_{\rm H}$=1.39$\times$10$^{23}$\,cm$^{-2}$, 
we obtain a much better fit with $\chi^2_{\rm red}$=0.8 (5 d.o.f.). 
As for the black-body model, letting $\Gamma$ or  $\cal{N}_{\rm H}$ free
does not lead to an improvement in the fits.
The corresponding absorption corrected flux in the 2--10\,keV range is
about 9.2 $\times$ 10$^{-14}$\,erg\,s$^{-1}$, implying a luminosity of 
about 10$^{33}$ d$^{2}_{10}$ erg\,s$^{-1}$.

\begin{figure*}[!Ht]
\hspace{-2.5cm}
\begin{tabular}{cc}
\hspace{3.5cm}\psfig{figure=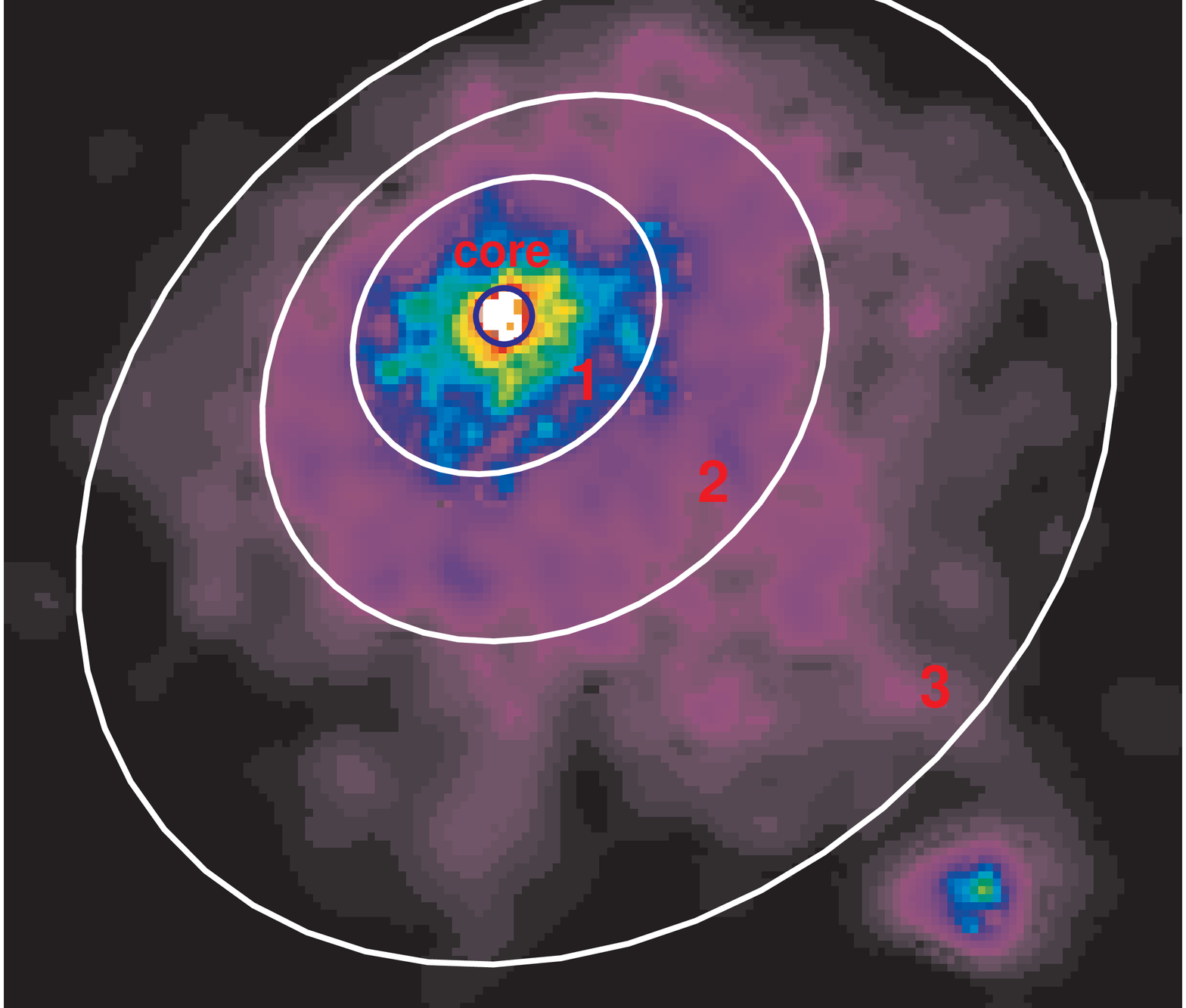,height=6cm,angle=-90} &
\psfig{figure=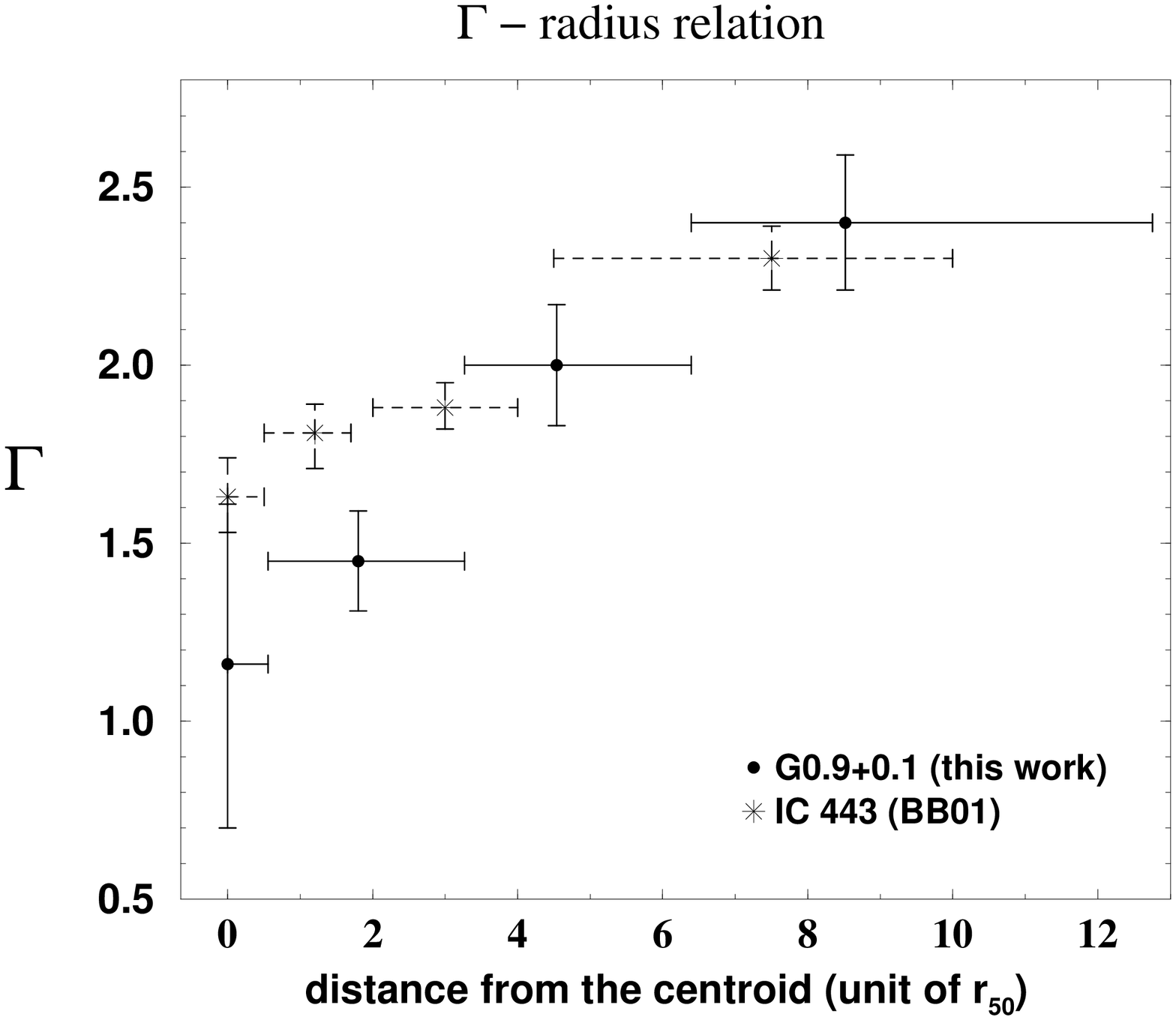,height=8.5cm,angle=-90}
\end{tabular}
\hspace{-2.5cm}
      \caption[]{{\it Left panel}: 
{\sl XMM-Newton} EPIC image (2.7\,\arcmin\,$\times$\,2.7\arcmin) 
of the PWN within G0.9+0.1, in the 3--8\,keV energy range, 
obtained with an adaptive smoothing filter with a signal-to-noise 
requirement of 5, and gaussian smoothing of 10$\arcsec$.
The regions used for the spectral
 analysis are superposed: core, regions 1, 2, and 3.
 {\it Right panel}: Photon index versus distance from the core.  
The X-axis (defined in Bocchino \& Bykov \cite{BocchinoBykov2001}) 
shows the weighted mean distance of the
 pixels of a given region from the centroid of the nebula, 
 expressed in units of r$_{50}$, the radius at
which the plerion surface brightness drops by a factor of 2 (7.15\arcsec~  for G0.9+0.1). 
 The r$_{50}$ unit gives a measure independent of the distance to
  the nebulae. For comparison, data for IC\,443 are also plotted 
(taken from Bocchino \& Bykov \cite{BocchinoBykov2001}). 
}
   \label{fig:PWN}
\end{figure*}
\section{Large-scale structure in the PWN: 
tests of the plerionic interpretation}\label{sec:plerion}

On large-scales the X-ray  morphology of the PWN determined from the 
{\sl XMM-Newton} observations 
is in relatively good agreement with the overall
radio emission.
The {\sl XMM-Newton} data 
enable us for the first time to study the variation of the spectral 
index within the PWN. In other plerions, a softening of the spectrum with
increasing radius has been observed; for example in 3C 58 
(Torii et al. \cite{Torii2000}, Bocchino et al. \cite{Bocchino2001}), 
G21.5-0.9 
(Slane et al. \cite{Slane2000}, Warwick et al. \cite{Warwick2001}) and 
IC 443 
(Bocchino \& Bykov \cite{BocchinoBykov2001}). This 
spectral softening can be explained 
by the shorter lifetime of high energy electrons compared to
lower energy 
electrons. 
To look for this effect, we extract the spectra 
in four regions, which are displayed in 
Figure~\ref{fig:PWN} ({\it left panel}).
The outer ellipse corresponds to the region taken 
for the spectral analysis of the overall PWN (see $\S$\ref{sec:PWN}). 
The observed spectra and best-fit models for  
regions 1 and 3  are shown, as examples, in Figure~2 in 
Porquet, Decourchelle \& Warwick (\cite{Porquet2002a}). 
  \begin{table}[!h]
\caption{Best-fit parameters for the combined (MOS+PN) spectrum of the 
large-scale structures of the PWN regions (defined in Fig.~\ref{fig:PWN}:
 {\it left panel}). 
 The model is an absorbed power law model ({\sc tbabs*po}). 
The cross-sections of the interstellar absorption are 
from Wilms et al. (2000). 
Unabsorbed X-ray fluxes (2--10\,keV) are expressed in 10$^{-12}$\,ergs\,cm$^{-2}$\,s$^{-1}$.  
{\it At the top}: ${\cal{N}}_{\rm H}$ is frozen to the value 
obtained for the entire PWN (see Table~\ref{table:table1}). 
{\it At the bottom}: ${\cal{N}}_{\rm H}$ is a free parameter.} 
\begin{tabular}{cccccc}
\hline
\hline
                 &  ${\cal{N}}_{\rm H}$ &  $\Gamma$ &  $\chi^{2}$/d.o.f.  &   F$^{\mathrm \tiny unabs}_{2-10}$   \\
           & ({\tiny $10^{23}$\,cm$^{-2}$})&               &              &\\
\hline
\hline
core       &1.39                             & 1.16$^{+0.45}_{-0.46}$  & {\small 18.2/18}   & 0.24\\
region 1   &1.39                             & 1.45$^{+0.14}_{-0.14}$  & {\small 105.7/146} &  1.80\\
region 2   &1.39                             & 2.00$^{+0.17}_{-0.17}$  & {\small 121.2/136} &  1.66\\
region 3   &1.39                             & 2.40$^{+0.19}_{-0.19}$  & {\small 162.2/173} & 2.02 \\
\hline
\hline
core       &1.36$^{+0.82}_{-0.61}$ & 1.12$^{+1.11}_{-0.88}$ & {\small 18.2/17}   & 0.24\\
region 1   &1.54$^{+0.25}_{-0.22}$  & 1.61$^{+0.30}_{-0.28}$ & {\small 104.5/145}&  1.95\\
region 2   &1.53$^{+0.27}_{-0.24}$  & 2.18$^{+0.37}_{-0.35}$ & {\small 120.2/135}&  1.83\\
region 3   &1.35$^{+0.24}_{-0.22}$  & 2.35$^{+0.39}_{-0.35}$ & {\small 162.1/172}&  1.96\\
\hline
\hline
&\\
\end{tabular}
\label{table:table2}
\end{table}
We fit the spectra of these regions with an absorbed power-law model
 ({\sc tbabs*po}), fixing $\cal{N}_{\rm H}$ to the value 
obtained for the overall PWN, 
i.e. $\cal{N}_{\rm H}\sim$ 1.4$\times$10$^{23}$\,cm$^{-2}$ 
(see Table~\ref{table:table1}). 
The corresponding best-fit parameters are reported in Table~\ref{table:table2}.
The value of the photon index $\Gamma$ shows a clear steepening of the 
spectrum from the inner part toward the outer part of the PWN 
(Fig.~\ref{fig:PWN}: {\it right panel}).
Freeing the absorption column density does not result in significant 
better fits.  
There is no evidence for any absorption variation across the PWN, 
as the column densities are compatible to within 12$\%$ for the different 
regions of the nebula. 
\begin{figure*}[!ht]
\includegraphics[angle=0,width=\textwidth]{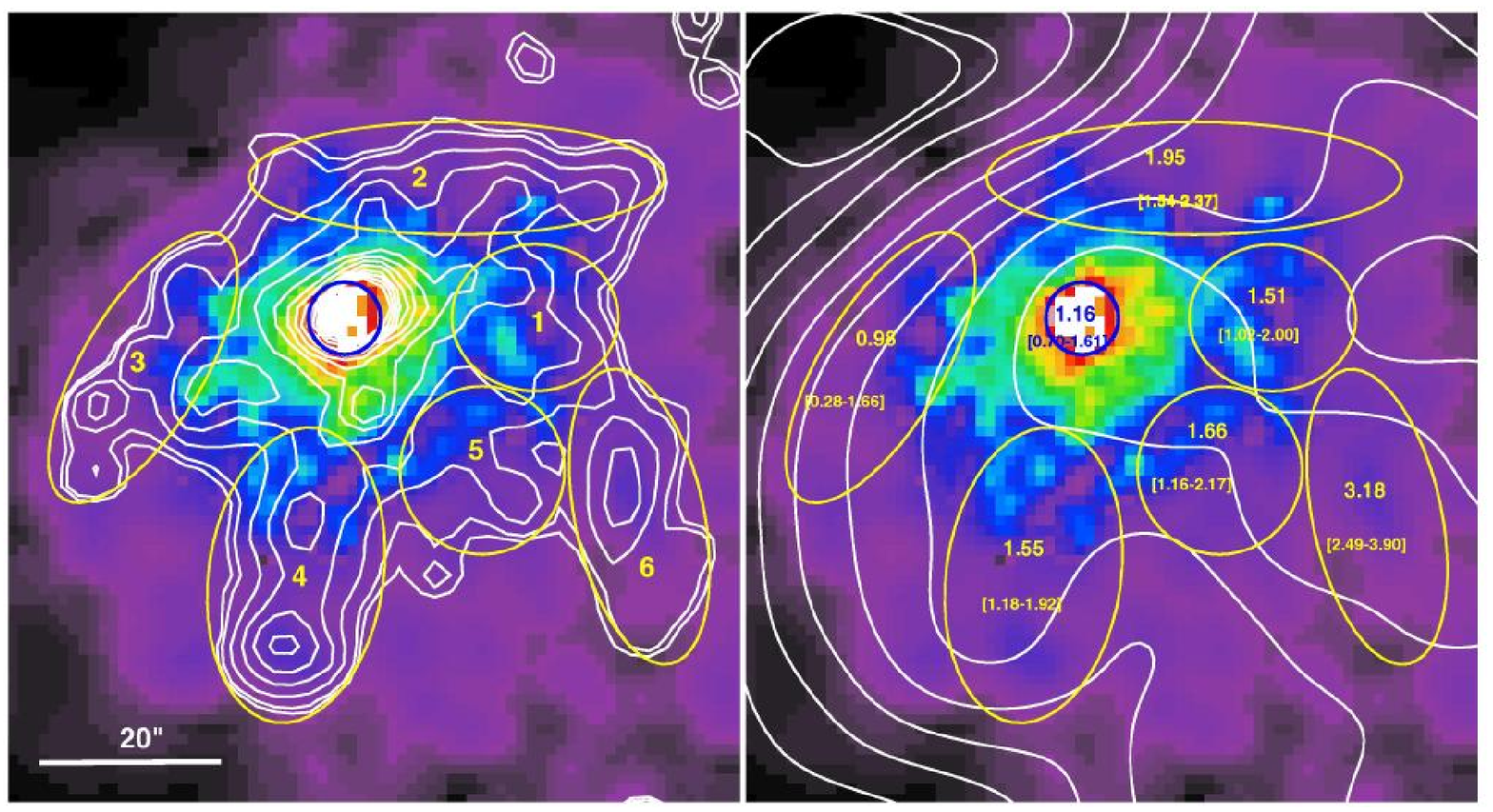}
\caption{Close-up view with XMM-Newton (EPIC) of the center of the PWN 
in the 3--8\,keV energy band obtained with an adaptative smoothing filter
with a signal-to-noise requirement of 5, and gaussian smoothing of 10$\arcsec$. 
{\it Left panel}: white contours correspond 
to the Chandra observation in the same energy range. 
 Yellow regions represent different small-scale structures surrounding the 
bright central core (dark blue circle) 
 matching the Chandra contours (see Gaensler et al. 2001):
 ``East arc-like feature'' (structure number 3),  
``jet-like feature'' (structure 4), and ``West arc-like feature'' (structure 6).
{\it Right panel}: the radio {\sl VLA} contours at 20\,cm overlaid
on the same XMM-Newton image.
The corresponding spectral indices  
and the minimum and maximum values (in brackets)
are given within 90$\%$ confidence uncertainties.}
\label{fig:smallstruct}
\end{figure*}

The spectral softening observed from the core to 
the outskirts of the PWN is consistent with synchrotron radiation losses
of high energy electrons as they diffuse through the nebula.
For comparison we show in Figure~\ref{fig:PWN} ({\it right panel})
 the radial variation of the spectral index 
measured for IC\,443 
(Bocchino \& Bykov \cite{BocchinoBykov2001}).
Our data show indication of a harder core and a 
stronger softening from the core towards the 
outer part of the G0.9+0.1 nebula, than is seen in
IC\,443, and also in other PWNe such as 3C\,58 and G21.5-0.9 
(see Fig.~5 in Bocchino \& Bykov \cite{BocchinoBykov2001}). 
This suggests the possibility of a relatively strong magnetic field 
existing in the G0.9+0.1  pulsar and/or its SNR environment. 
The variation in the spectral index between the
core and the PWN  is roughly consistent
with the empirical linear relationship derived by 
Gotthelf \& Olbert (\cite{Gotthelf2001}) for 6 pulsars and
their respective PWNe, i.e. $\Gamma_{\rm PWN}$=0.8$\times 
\Gamma_{\rm core}$+0.8. SNR G54.1+0.3 shows a similar spectral index, 
$\Gamma$=1.09$^{+0.08}_{-0.09}$, for its pulsar region 
(Lu et al. \cite{Lu2002}), and $\Gamma$=1.9$\pm$0.2 for the overall PWN  
(Lu et al. \cite{Lu2001}). 

\section{Small-scale structures in the PWN}\label{sec:smallscale}

The high angular resolution of {\sl Chandra} has provided
an unprecedented view of the X-ray morphology of this PWN.
Relatively  small-scale features identified in the {\sl Chandra}
observations  (Gaensler et al. \cite{Gaensler2001})  include a semi-circular 
arc-like feature, a jet-like feature, 
and a bright clump close to a very faint unresolved source 
(CXOU J174722.8-280915, which as noted earlier is inferred to be 
the pulsar itself).
This X-ray morphology is similar to other PWNe   
powered by young pulsars which display jets combined with
a torus structure (e.g. Crab: Brinkmann et al. \cite{Brinkmann85}, 
Weisskopf et al. \cite{Weisskopf2000}; 
Vela: Pavlov et al. \cite{Pavlov2000}).
The X-ray brightest region matches the eastern radio peak 
while the western side of the X-ray arc-like feature 
(structure 6 in Fig.~\ref{fig:smallstruct}) corresponds to the western radio peak.
\begin{figure}[!hb]
\includegraphics*[height=7.5cm]{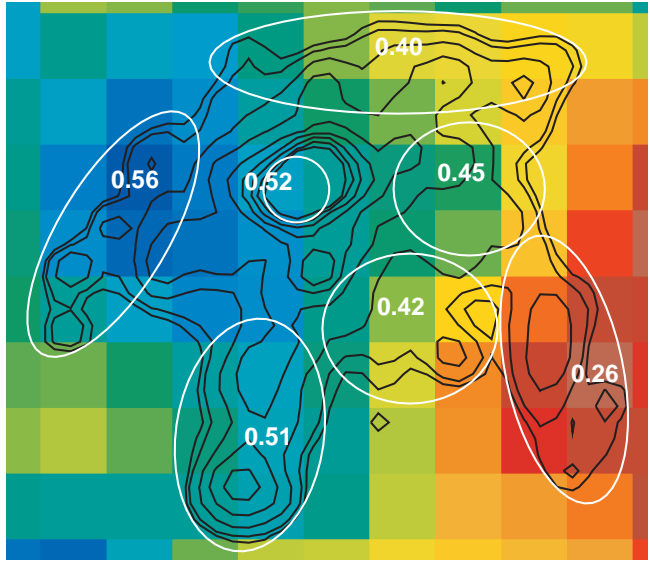}
\caption{Subtracted background hardness ratio map ([6--10]keV/[3--6]keV) 
of the small-scale structures with a pixel size of 8''. 
The colour coding goes from blue (hard spectrum) to red (soft
spectrum). To guide the eye we give explicit hardness ratio values for
characteristic pixels.
As in Figure~\ref{fig:smallstruct} ({\it left panel}), 
 the Chandra contours and the regions taken for our spectral analysis are superposed.}
\label{fig:HR}
\end{figure}
In the radio band, as displayed in Figure~1 in 
Helfand \& Becker (\cite{Helfand87}), 
there is a spectral index change between 20\,cm and 6\,cm, i.e.
the eastern peak present at 20\,cm disappears at 6\,cm.
 The jet-like feature  (structure 4 in Fig.~\ref{fig:smallstruct}) 
corresponds to a distortion in the radio contours. \\
\begin{table}[!h]
\caption{Best-fit parameters for the combined (MOS+PN) spectrum 
of the small-scale structures of the PWN of G0.9+0.1 
(defined in Fig.~\ref{fig:smallstruct}).
 The model is an absorbed power law model ({\sc tbabs*po}).  
Unabsorbed X-ray fluxes (2--10\,keV) are expressed in 
10$^{-12}$\,ergs\,cm$^{-2}$\,s$^{-1}$.  ${\cal{N}}_{\rm H}$ 
is fixed at 1.39$\times$10$^{23}$\,cm$^{-2}$.}
\begin{tabular}{cccccc}
\hline
\hline
structure   &     $\Gamma$                &  $\chi^{2}$/d.o.f.  &   F$_{\mathrm X}^{(b)}$   & {\sl Chandra}\\
                   &                                       &                              &{\tiny (2--10\,keV)} &identification\\
\hline
\hline
core             & 1.16$^{+0.45}_{-0.46}$   & {\small 18.2/18} &  0.24     & \\
    1            & 1.51$^{+0.49}_{-0.49}$   & {\small 13.5/16} &  0.22     & \\
    2            & 1.95$^{+0.42}_{-0.41}$   & {\small 21.8/25} &  0.32     &\\
    3            & 0.98$^{+0.68}_{-0.70}$   & {\small 15.9/14} &  0.20     &  ``East arc''? \\
    4            & 1.77$^{+0.41}_{-0.41}$   & {\small 13.6/21} & 0.28      &  ``jet'' ?\\
    5            & 1.66$^{+0.51}_{-0.50}$   & {\small 19.1/18} & 0.23      &\\        
    6            & 3.18$^{+0.72}_{-0.69}$   & {\small 13.8/12} & 0.22      & ``West arc''?\\
\hline
\hline
&\\
\end{tabular}
\label{table:jetarc}
\end{table}
 In our {\sl XMM-Newton} close-up view, these small-scale features  
are not as clearly detected as in the {\sl Chandra} data. However,
the extent of this emission is about a factor 2 larger than observed with
 {\sl Chandra}, thanks to the high sensitivity of {\sl XMM-Newton}. 
Using as guide the locations defined by {\sl Chandra},
 we are able for the first time to derive the spectra of the various
small-scale structures seen in  the G0.9+0.1 PWN.
 Determining the spectral variations in these small-scale structures 
 can give an indication of the geometry and orientation of the nebula.
The best-fit parameters of the combined fits (MOS and PN)  
of the  six regions identified in Figure~\ref{fig:smallstruct}
are given in Table~\ref{table:jetarc}.  Figure~\ref{fig:smallstruct} 
({\it right panel}) also displays the spectral index measured 
for each region. 
The region corresponding to the eastern part of the arc-like feature, which
was first revealed by {\sl Chandra}, shows apart from the 
core region ($\Gamma$=1.2$\pm$ 0.5)
the hardest spectrum with a spectral index of ($\Gamma$=1.0$\pm$0.7) 
among all the structures of the PWN. 
The former region encloses some unresolved bright knots which may be
point-like sources, however if we exclude these bright sources 
we still obtain a very flat power-law form with
 $\Gamma\sim$ 1.1$\pm$0.8 ($\chi^{2}_{\rm red}$=1.2, 12 d.o.f). 
In contrast the region corresponding to the western side  
 of the arc-like feature, appears to 
have the steepest photon index ($\Gamma\sim$3.2$\pm$0.7)
of any region within the PWN. 
In order to investigate spectral variations of the small
scale structures, we constructed a subtracted background hardness ratio image 
([6--10]keV/[3--6]keV) of the central region (see Fig.~\ref{fig:HR}). 
It can be clearly seen that the eastern part of the
arc-like feature shows harder X-ray emission than the western part. 
 A possible explanation is that the east arc-like feature is 
pointing towards the observer, and that its spectral hardness is due to the 
relativistic beaming or Doppler boosting of the electrons, while the 
western arc corresponds to the opposite part of the torus suggested 
in the {\sl Chandra} data. 
 The region associated with the jet-like feature 
in Gaensler et al. (\cite{Gaensler2001}) does not exhibit a harder spectrum as 
suggested by their inferred hardness ratio. 
 Splitting this structure into two parts, north and south, 
and fitting the spectra separately gives respectively 
 as spectral spectral indexes $\Gamma \sim $~1.4$\pm$0.9,  
and $\Gamma \sim $~2.1$\pm$0.5. 
The possible observed curvature of this 
putative jet from small-scale to large-scale may be the result of an
interaction with the high magnetic fields of the Galactic Center
region.

\section{Summary}
\vspace*{0.1mm}
\indent We present first results of an {\sl XMM-Newton} observation of the
 composite supernova remnant G0.9+0.1 located in the Galactic Center 
region. \\
\indent The high sensitivity of the {\sl XMM-Newton}
 observatory allows for the first time 
to detect diffuse X-ray emission in the region of the radio shell. We find an 
unabsorbed flux of about 2$\times$10$^{-12}$ erg\,cm$^{-2}$\,s$^{-1}$ (2--10\,keV). \\
\indent  The {\sl XMM-Newton} X-ray morphology is in relatively 
good agreement with published ({\sl VLA}) radio maps of the PWN.
 We have carried out the first detailed X-ray spectral analysis of the
PWN inside G0.9+0.1.  On large scales, there is a clear
softening of the spectrum with radial distance from the core, 
a phenomenon which has been observed previously in a number
of X-ray plerions (e.g., 3C58, G21.5-0.9, IC443).
 Detailed spectral analysis is also presented of the small-scale 
structures evident in the PWN.  
The eastern part of the arc-like 
feature (revealed by {\sl Chandra}) presents a reliable indication  
for a very hard photon index ($\Gamma \sim$1.0$\pm$0.7), 
while the western part presents a very soft spectrum ($\Gamma \sim$3.2$\pm$0.7).
 A possible explanation is that the east arc-like feature is 
pointing towards the observer, and that its spectral hardness is due to the 
relativistic beaming or Doppler boosting of the electrons, while the 
western arc corresponds to the opposite part of the torus suggested 
in the {\sl Chandra} data. 

\indent G0.9+0.1 provides further 
clues concerning the processes by which pulsars connect with their 
environment and is also important to illustrate a possible impact of the strong 
external magnetic field that pervades the Galactic Center region.

\begin{acknowledgements}
This work is based on observations obtained with {\sl XMM-Newton}, 
an ESA science mission with instruments and contributions directly funded by ESA Member
 States and the USA (NASA).  
The authors would like to acknowledge B.M. Gaensler for
providing the radio {\sl VLA} image. 
We would like to thank Jean Ballet for a careful reading of the manuscript, 
as well as the anonymous referee for valuable comments and suggestions.
D.P. would like to thank the
X-ray team at Saclay for developing much of the software used in
the present analysis.
D.P. thanks D.M. Neumann for fruitful discussions about data reductions 
and statistics. 
D.P. acknowledges grant support from the 
``Institut National des Sciences de l'Univers'' and from the
``Centre National d'Etudes Spatial''.
\end{acknowledgements}


\begin{thebibliography}{}
\bibitem[1989]{Anders89} 
Anders E., Grevesse N., 1989, Geochimica et Cosmochimica Acta,  53, 197
\bibitem[2001]{Arnaud2001} 
Arnaud M., Neumann D.~M., Aghanim N., Gastaud R., 
Majerowicz S., Hughes J.~P., 2001, A\&A,  365, L80 
\bibitem[1997]{Becker97} 
Becker W., Truemper J., 1997, A\&A,  326, 682. 
\bibitem[2001]{BocchinoBykov2001} 
Bocchino F., Bykov A.~M., 2001, A\&A,  376, 248
\bibitem[2001]{Bocchino2001} 
Bocchino F., Warwick R.~S., Marty P., Lumb D., 
Becker W., Pigot C.  2001, A\&A,  369, 1078
\bibitem[1985]{Brinkmann85}
Brinkmann W., Aschenbach B., Langmeier A. 1985, Nature, 313, 662
\bibitem[2001]{Gaensler2001}
Gaensler B.M., Pivovaroff M.J., Garmire G.P. 2001, ApJ, 556, L107
\bibitem[2001]{Gotthelf2001}
Gotthelf E.V. \& Olbert C.M., 2001, in ``Neutron Stars in Supernova Remnants'' 
(ASP Conference Proceedings), eds P. O. Slane and B. M. Gaensler, 
in press [astro-ph/0112017]
\bibitem[2001]{Green2001}
Green D.A., 2001, ``A Catalogue of Galactic Supernova Remnants'', 
Mullard Radio Astronomy Observatory, Cavendish Laboratory, 
Cambridge, United Kingdom (http://www.mrao.cam.ac.uk/surveys/snrs/). 
\bibitem[1987]{Helfand87}
Helfand D.J. \& Becker R.H. 1987, ApJ 314, 203
\bibitem[2001]{Lu2001} 
Lu F.~J., Aschenbach B., Song L.~M., 2001, A\&A,  370, 570
\bibitem[2002]{Lu2002}
Lu F.J., Wang Q.D., Aschenbach B., Durouchoux Ph., Song L.M. 2002, ApJ, 568, L49
\bibitem[2002]{Maj2002} 
Majerowicz S., Neumann D.~M., Reiprich T.~H., 2002, A\&A, 394, 77 
\bibitem[1998]{Mereghetti98} 
Mereghetti S., Sidoli L., Israel G. L. 1998, A\&A, 331, L77
\bibitem[1983]{Morrison83} 
Morrison R. \& McCammon D., 1983, ApJ,  270, 119
\bibitem[2000]{Pavlov2000}
Pavlov G.~G. et al. 2000, AAS, 196, 3704
\bibitem[2002]{Porquet2002a}
Porquet D., Decourchelle A., Warwick R.S. 2002, proceedings of the symposium 'New Visions of the X-ray Universe in the XMM-Newton and Chandra Era', Ed. F. Jansen, in press [astro-ph/0204261]
\bibitem[2000]{Sidoli2000} 
Sidoli L., Mereghetti S., Israel G.\ L., Bocchino F. 2000, A\&A, 361, 719 
\bibitem[2000]{Slane2000} 
Slane P., Chen Y., Schulz N.~S., Seward F.~D., Hughes J.~P., Gaensler B.~M 2000, ApJ,  533, L29
\bibitem[2000]{Torii2000} 
Torii K., Slane P.~O., Kinugasa K., Hashimotodani K., Tsunemi H. 2000, PASJ,  52, 875
\bibitem[1995]{vanRiper95} 
van Riper K.~A., Link B., Epstein R.~I., 1995, ApJ,  448, 294. 
\bibitem[2001]{Warwick2001} 
Warwick R.~S., Bernard J.-P., Bocchino F., et al. 2001, A\&A,  365, L248
\bibitem[2000]{Weisskopf2000} 
Weisskopf M.~C., Hester J.~J., Tennant A.~F., et al., 2000, ApJ,  536, L81
\bibitem[2000]{Wilms2000} 
Wilms J., Allen A., McCray R., 2000, ApJ,  542, 914
\end{thebibliography}
           \end{document}